\pdfoutput=1
\documentclass[12pt]{article}
\usepackage[margin=1in]{geometry}

\usepackage{amsmath,dsfont}
\usepackage{amssymb}
\usepackage{graphicx,color}
\usepackage{slashed,cite}

\usepackage{ifpdf}

\newcommand{\be}{\begin{equation}}
\newcommand{\ee}{\end{equation}}
\newcommand{\ba}{\begin{eqnarray}}
\newcommand{\ea}{\end{eqnarray}}

\title{{\bf \Large{The dS swampland conjecture and the Higgs potential}}\bigskip}

\author{ Frederik Denef,$^a$ Arthur Hebecker,$^b$ and Timm Wrase$^c$\bigskip\bigskip\\
\small $^a$Department of Physics, Columbia University\\ \small 538 West 120th Street, New York, NY 10027, USA\bigskip\\
\small $^b$Institute for Theoretical Physics, University of Heidelberg, \\ \small Philosophenweg 19, D-69120 Heidelberg, Germany\bigskip\\
\small $^c$Institute for Theoretical Physics, TU Wien,\\ \small Wiedner Hauptstrasse 8-10/136, A-1040 Vienna, Austria\bigskip
}

\date{}

\begin{document}

\maketitle

\abstract{
\noindent
According to a conjecture recently put forward  in \cite{Obied:2018sgi}, the scalar potential $V$ of any consistent theory of quantum gravity satisfies a bound $|\nabla V|/V \geq {\cal O}(1)$. This forbids dS solutions and supports quintessence models of cosmic acceleration. Here we point out that in the simplest models incorporating the Standard Model in addition to quintessence, with the two sectors decoupled as suggested by observations, the proposed bound is violated by 50 orders of magnitude. However, a very specific coupling between quintessence and just the Higgs sector may still be allowed and consistent with the conjecture.
}

\newpage

\section{Introduction}
Compactifications of superstring or M-theory to 4d Minkowski or AdS space are well understood and have led to the idea that string theory gives rise to a gigantic landscape of possible 4d low energy effective theories \cite{Strominger:1986uh,Dasgupta:1999ss,Bousso:2000xa,Giddings:2001yu,Kachru:2003aw,Susskind:2003kw, Douglas:2003um,Denef:2004ze,Grana:2005jc,Douglas:2006es}. However, classical no-go theorems such as \cite{Maldacena:2000mw} indicate that realizing {\it de Sitter} vacua in string theory requires quantum and/or stringy ingredients. The fact that corrections to classical 10d low energy supergravity are qualitatively important implies that dS compactifications, in contrast to AdS or Minkowski compactifictions, must live in a regime in which these corrections cannot be made arbitrarily small \cite{Dine:1985he}, hence perturbation theory cannot be made arbitrarily accurate. Moreover the absence of supersymmetry in dS, and perhaps more fundamentally the lack of a complete, nonperturbative formulation of string theory, make it hard to obtain exact results beyond perturbation theory. Thus a completely rigorous, parametrically controlled construction of individual de Sitter vacua in string theory has remained out of reach. On the other hand, starting with \cite{Kachru:2003aw}, much progress has been made over the past 15 years in developing models containing all the ingredients needed to produce effective potentials generic enough to support an abundance of dS vacua, barring extraordinary conspiracies that would somehow eliminate all of those.\footnote{See for example \cite{Kachru:2003sx,Burgess:2003ic,Saltman:2004sn,Denef:2004dm,Denef:2004cf,Saltman:2004jh,Balasubramanian:2005zx,Westphal:2006tn} for a sample of papers from just the first three years, \cite{Aalsma:2018pll, Gallego:2017dvd, Buchmuller:2016dai, Retolaza:2015nvh, Cicoli:2015ylx, Achucarro:2015kja, Braun:2015pza} for a sample of papers from just the last three years, and \cite{Denef:2008wq,Silverstein:2016ggb, Baumann:2014nda} for many more references. See \cite{Danielsson:2018ztv} for an overview of    conspiracies rendering constructions of dS vacua more challenging than one might naively expect.}

It is nevertheless interesting to entertain the possibility that what looks like an extraordinary conspiracy from a low energy effective field theory point of view might in fact be the consequence of a simple fundamental property of quantum gravity. Along these lines, the authors of \cite{Obied:2018sgi} have put forward the audacious conjecture that the low energy effective scalar potential $V$ in any consistent theory of quantum gravity must satisfy 
\begin{equation} \label{OOSV}
 \frac{|\nabla V|}{V} \geq c \, , \qquad c \sim {\cal O}(1) \, .
\end{equation}
Here $|\nabla V|$ is the norm of the gradient of $V$ on the scalar manifold, $c$ is a constant of order unity, and the reduced Planck mass has been set to  $M_P=1$. This has been further studied and scrutinized in \cite{Agrawal:2018own, Agrawal:2018mkd, Andriot:2018wzk, Banerjee:2018qey, Aalsma:2018pll, Achucarro:2018vey, Garg:2018reu, Lehners:2018vgi, Kehagias:2018uem}. 

The bound in equation (\ref{OOSV}) implies in particular the absence of metastable de Sitter vacua, but is much stronger than that. Note in particular that it could be falsified by finding {\it any} positive-$V$ critical point, even if it is just a local maximum or saddle point. At first glance it seems that such critical points were constructed in e.g. \cite{Caviezel:2008tf, Flauger:2008ad, Danielsson:2011au} in a class of models that was actually used to motivate the dS swampland conjecture of \cite{Obied:2018sgi}. However, as was shown in \cite{Danielsson:2011au}, imposing proper flux quantization in the simplest example forces these dS critical points to be at small volume and large string coupling. 

While it would be interesting to establish the existence of such a positive-energy hilltop potential in string theory, we want to take a bottom-up approach in this note. Our starting point is the trivial observation that the Standard Model Higgs potential provides such a maximum at zero Higgs VEV. Coupling to other fields such as a quintessence scalar may reinstate a nonvanishing $|\nabla V|$ at this point. However, tight observational bounds constrain the interaction of extra light fields with the Standard Model. In the simplest decoupled quintessence + Higgs model, we find $|\nabla V|/V \sim 10^{-55}$, in considerable tension with the assertion that $c$ in equation (\ref{OOSV}) is of order unity. A special coupling of the quintessence field to only the Higgs may restore $c \sim {\cal O}(1)$, although this appears unnatural, at least from a 4d effective field theory point of view. More work is required to determine whether this indicates a fatal flaw or a powerful prediction of equation (\ref{OOSV}), as we outline in the discussion section.

\section{The Higgs potential and Quintessence}
Concretely, the perturbative Standard Model Higgs potential 
\be
V_H=\lambda_H(|H|^2-v^2)^2
\ee
has an (SU(2)-symmetric) maximum at $H=0$. The height difference $\Delta V_H$ between the minimum and this maximum is 
\be
\Delta V_H=\lambda_Hv^4 \sim m_H^2v^2 \sim (125\,\mbox{GeV})^2 (250\,\mbox{GeV})^2\sim 10^{-65} M_P^4\,.
\ee

Now, as advocated in \cite{Obied:2018sgi,Agrawal:2018own}, let us ascribe the present cosmic acceleration to a rolling quintessence scalar \cite{Wetterich:1987fm, Ratra:1987rm, Caldwell:1997ii} rather than to a cosmological constant (the latter being inconsistent with equation (\ref{OOSV})). This in principle allows one to avoid $c\ll1$. Concretely, it was argued in \cite{Agrawal:2018own} that current observational constraints on dark energy only require $c \lesssim .6$, consistent with the proposed swampland criterion, and that the least constrained model is of the form
\be
V_Q(\phi) = V_0 \, e^{-\lambda \phi}\,.
\ee
Hence, $\lambda\lesssim 0.6$ and $|\nabla V_Q|_{\rm today}=\lambda V_Q(\phi_{\rm today})\sim 10^{-120}$ in Planck units. The property $|\nabla V_Q| \sim V_Q \sim 10^{-120}$ is a generic feature of quintessence models for the currently observed cosmic acceleration.

Now, combining this additively with the Higgs potential,
\be
V=V_Q(\phi)+V_H(H)\,,\label{vsum}
\ee
we have an EFT description of our world which should obey the conjecture in equation (\ref{OOSV}) at any point of its field space. However, at the present value of $\phi$ and $H=0$, we have
\be
\frac{|\nabla V|}{V} \simeq \frac{|\nabla V_Q|_{today}}{\Delta V_H} \sim \frac{10^{-120}}{10^{-65}}\sim 10^{-55}
\ee
in Planck units, evidently in serious tension with the conjecture. 

One could try to avoid the problem by giving a large gradient to the symmetric point of the Standard Model Higgs potential. As is well known, and as was emphasized also in section 3 of \cite{Agrawal:2018own} and earlier in \cite{Brennan:2017rbf}, the quintessence field cannot be substantially coupled to visible matter, in order to be consistent with the observed absence of time variation of Standard Model parameters such as the fine-structure constant, and with tests of the equivalence principle, i.e.\ the absence of a fifth force. Nevertheless, certain appropriately chosen but from an EFT point of view extremely fine-tuned combinations of couplings to just the Higgs, but not to other Standard Model fields, may restore equation (\ref{OOSV}). A particularly simple possibility is
\be \label{Hphicoupling}
V(\phi,H)=e^{-\lambda\phi}\,(V_H(H)+\Lambda)\,,
\ee
where $\phi_{\rm today}=0$ and $\Lambda$ corresponds to today's dark energy density. This Lagrangian induces a trilinear coupling between the quintessence scalar $\phi$ and the physical Higgs field $h$ (schematically, $H=v+h$):
\be
{\cal L}\supset \phi h^2\,\frac{v^2}{M_P}\,,\label{hqc}
\ee
where we have reinstated the Planck mass explicitly for convenience of the phenomenological discussion. 

The quintessence field necessarily has a tiny effective mass $m \lesssim H_0 \sim 10^{-33} \, {\rm eV}$ \cite{Tsujikawa:2013fta}. Tight upper bounds exist on the coupling of such light scalars to the Standard Model, from a variety of observations. We will focus here on high-precision tests of the equivalence principle, summarized in \cite{Wagner:2012ui}. For Yukawa interactions of the form
\begin{equation} \label{Yuk}
 {\cal L}_{\rm Yuk} = g \, \phi N N \, ,
\end{equation}
where $N$ is a nucleon field, $g$ is the dimensionless Yukawa coupling, and $\phi$ is a scalar with mass $m < 10^{-13} \, {\rm eV}$, the bound reported in \cite{Wagner:2012ui} is
\begin{equation} \label{expbound}
 g < 10^{-24} \, . 
\end{equation}
In the model under consideration, we have declared the quintessence field $\phi$ to be exactly decoupled from all fundamental Standard Model fields except the Higgs. However the coupling in equation (\ref{hqc}) will still induce an effective coupling of the form in equation (\ref{Yuk}) through loop diagrams. Most naively, one needs a Higgs loop and two-light-quark Yukawas to induce a coupling $\phi NN$ with two nucleons. This coupling will be suppressed at least by the factor $v^2/M_P$ from equation (\ref{hqc}) and, additionally, by two light-quark Yukawas $y_d^2\sim 10^{-10}$ as well as the nucleon-mass/Higgs-mass ratio $\sim m_p/v^2$. The mass dimension $-1$ of this factor follows from dimensional analysis. The appearance of (at least one power of) a soft scale in the numerator is required since, without quark momenta or gluon vertices, the relevant diagram simply gives zero. Thus, in total we expect
\be \label{bound1}
{\cal L}\supset Y\,\phi NN\qquad \mbox{with}\qquad Y\lesssim \frac{m_py_d^2}{M_P}\,\frac{1}{16\pi^2}\sim 10^{-30}\,.
\ee
Alternatively, and perhaps more importantly, one can attach the two Higgs lines to a $Z$ boson line and then connect the latter to a quark line. This is a two-loop loop diagram in which one must furthermore use a light-quark mass in the fermion propagator between the two $Z$ vertices. This is necessary to introduce a sensitivity to the type of light quark, such that an equivalence-principle-violating effect results. Thus, one now has only one Yukawa, at the price of an additional electroweak loop suppression factor $g^4/(16\pi^2)\sim 10^{-3}$:
\be \label{bound2}
{\cal L}\supset Y'\,\phi NN\qquad \mbox{with}\qquad Y'\lesssim \frac{m_py_d}{M_P}\,\frac{g^4}{(16\pi^2)^2}\sim 10^{-28}\,.
\ee
Both estimates in equations (\ref{bound1}) and (\ref{bound2}) are well below the bound in equation (\ref{expbound}). Thus we conclude that tests of the equivalence principle cannot at present exclude the specific quintessence model under consideration. Of course this model, or variants thereof, will have to satisfy many additional consistency checks, like for example from reheating. It would be interesting to find out whether one can exclude the above model based on additional observational constraints.

The crucial assumption in the setup above is that the quintessence field is coupled to the Higgs exactly as in equation (\ref{Hphicoupling}), and decoupled from all other Standard Model fields. Clearly this is highly fine-tuned from a pure 4d effective field theory point of view. Indeed, even disregarding the decoupling from all other fields, we have in equation (\ref{Hphicoupling}) demanded an identical coupling of the quintessence scalar to the quartic Higgs term, to the Higgs mass term and to the constant $\sim v^4$, which in general all renormalize independently. Without this identical coupling, we would have obtained a dependence of the Higgs VEV on $\phi$, a mixing of $\phi$ with the Standard Model Higgs, and stronger bounds (in particular from the induced cosmological evolution of the Higgs VEV). This situation is similar to that of an extra, super-light Higgs mixing with the Standard Higgs \cite{Ahlers:2008qc}. There, the mixing parameter between the super-light and the Standard Model Higgs was constrained to be below $10^{-21}$. 

On the other hand, in a string theory context, couplings like the one in equation (\ref{Hphicoupling}) may be more natural. Indeed the canonically normalized dilaton has such universal exponential couplings, as do various other scalars arising in string compactifications. However in general these scalars couple similarly to fermions, making it difficult to imagine how the coupling (\ref{Hphicoupling}) could naturally arise in this way while at the same time keeping the couplings of $\phi$ to the Standard Model fermions sufficiently suppressed to ensure the observational bound (\ref{expbound}).

Of course the swampland conjectures do not necessarily require the low energy effective field theory of the real world to satisfy any of our preconceived notions of naturalness, and it is possible that models looking tuned or contrived from an EFT point of view somehow naturally emerge from string theory or other consistent theories of quantum gravity. A thorough analysis of explicit string theory constructions would be needed to be conclusive either way. We leave this to future work.

Finally, there are certainly other ways of how on could try to remove the maximum in the Standard Model Higgs potential. For example, one might think that any modulus $\chi$, which is in its minimum in the Standard Model broken phase, will not in general be in its minimum in the symmetric phase, naturally providing a gradient. However, for sufficiently strongly stabilized moduli a slightly changed new minimum corresponding to the symmetric phase will exist, again providing a problematic $|\nabla V|=0$ locus. If $\chi$ is not sufficiently strongly stabilized, building such a model still involves major challenges. Indeed, replace $V_H(H)$ in equation (\ref{vsum}) by $V_H(H,\chi)$ and assume that $V_H(v,\chi)$ has a local minimum with value zero at $\chi=\chi_0$. Furthermore, consider the Higgs potential in the symmetric phase, $V_H(H=0,\chi)$, viewed as a function of $\chi$ alone. If the latter has a critical point above zero or asymptotes to a positive value, the conjecture will be violated. If it takes a negative value, one also encounters a problem: Indeed, one then has trajectories in $(H,\chi)$-field space connecting todays vacuum, $(H=v,\chi=\chi_0)$, with a point $(H=0,\chi=\chi_1)$, where $V_H(H=0,\chi_1)<0$. The lowest of these trajectories has a saddle at positive potential, violating the conjecture. It is possible to avoid this problem by having $V_H(v,\chi_0)<0$ (the order-of-magnitude should however not exceed that of today's dark energy density). In this case the whole trajectory must be very flat -- also a potentially difficult situation. Finally, the $(H,\chi)$-field space may have boundaries at finite distance and finite value of $V_H$, avoiding the problems described above. It remains to be seen whether such loopholes arise in actual string compactification scenarios compatible with known constraints on physics beyond the Standard Model. Again, we leave this to future work.

\section{Discussion}

We pointed out that the recently proposed dS swampland conjecture in equation (\ref{OOSV}) is violated by the Standard Model supplemented with a decoupled quintessence scalar. Our analysis left open the possibility that coupling the quintessence scalar to the Higgs as in equation (\ref{Hphicoupling}), while suppressing couplings to all other Standard Model fields so as to ensure equation (\ref{expbound}), is compatible both with equation (\ref{OOSV}) and with the absence of detectable long-range scalar forces. However this appears extremely fine-tuned, at least from an effective field theory point of view.

Although we did not study this here, similar arguments can presumably be made in the QCD sector of the Standard Model, to which the bounds on long-range scalar forces apply more directly. Indeed, the perturbative QCD vacuum has a higher energy than the confined phase with chiral symmetry breaking in which we live. Naturally, one would assume that an effective description with a tachyonic composite degree of freedom exists and that, starting in the perturbative QCD vacuum, the condensation of this degree of freedom would take us to the phase with broken chiral symmetry. The change in the potential energy is $\sim \Lambda_{QCD}^4$,  which is smaller by roughly 12 orders of magnitude compared to $\lambda_Hv^4$ in the Higgs case. This is of course still sufficient to exclude a $c$ of order unity. It would be interesting to explore this further.

A natural question is whether the decoupled Higgs + quintessence model is consistent with recent speculations on modifications of the conjecture. For example in \cite{Andriot:2018wzk, Garg:2018reu} it was proposed to allow small or vanishing first derivatives of the potential in regions where the mass matrix has large negative eigenvalues. This is consistent with the Higgs potential where at the maximum we have a large $\eta$ slow-roll parameter.\footnote{The same holds for dS critical points constructed in classical type II flux compactifications~\cite{Caviezel:2008tf, Flauger:2008ad}.} 
So with this note we have not excluded a ``quasi-de Sitter conjecture'' that would prevent small $\epsilon$ and $\eta$ slow-roll parameters for any point in moduli space with $V>0$. Relatedly, a milder conjecture was formulated in \cite{Dvali:2018fqu} (based on earlier work \cite{Dvali:2013eja, Dvali:2014gua, Dvali:2017eba}) for which $c$ did not have to be constant but could depend on the value of the scalar potential. The given absolute lower bound of $c=V$ is not violated by the decoupled Higgs + quintessence model.

In addition to testing its consequences, it is imperative to understand how  the dS swampland conjecture relates to concrete string models. Indeed, most constructions of metastable de Sitter vacua are based on moduli stabilization (in AdS) and uplifting due to a SUSY-breaking sector. Part of the uplift energy can be ascribed to loops of the non-SUSY Standard Model -- an effect much larger than the observed dark energy density. Thus, in any case, a fine compensation between positive and negative energy terms is required. This would still be true if our world was (very weakly) AdS. Now, if all those moduli-stabilization/uplifting constructions were flawed due to some overlooked instability or inconsistency of the 4d EFT approach, then one would not only lose dS models but plausibly also all metastable non-SUSY AdS models with small $|\Lambda|$. It is furthermore not obvious why a quintessence scalar with practically flat potential after SUSY breaking, a feature not generally present in the string constructions mentioned above, would help. This certainly calls for more work.

To conclude, one can interpret our results either as suggesting that the dS swampland conjecture as stated in equation \eqref{OOSV} is not correct, or as suggesting that the conjecture is even more powerful than previously thought, possibly leading to experimentally testable universal predictions, such as long-range scalar forces. In order to settle this, the following two concrete key questions need to be addressed:
\begin{itemize}
 \item What are the constraints on general 4d (super)gravity theories incorporating the Standard Model + quintessence, such that the theoretical bound in equation (\ref{OOSV}) is satisfied as well as observational bounds such as the one in equation (\ref{expbound})? 
 \item $\!$Do suitable explicit string theory realizations of the Standard Model + quintessence  satisfy these constraints without excessive fine tuning?  
\end{itemize} 
We hope this note will serve as encouragement to make progress along these lines and to ultimately answer these two questions.

\vspace*{.2cm}
\noindent
{\bf Acknowledgements:} 
We are particularly grateful to D.~Andriot, T.~Bachlechner, M.~Bauer, W.~Buchm\"uller, L.~Hui, J.~Jaeckel, S.~Kachru, R.~Kallosh, S.~Leonhardt, A.~Linde, T.~Plehn, M.~Ratz, P.~Soler and the participants of the Bethe Forum ``String Theory Challenges in Particle Physics and Cosmology'' for enlightening discussions. TW is supported by an FWF grant with the number P 30265. We acknowledge the support of the Bethe Center for Theoretical Physics, Bonn. FD is supported in part by the US Department of Energy grant de-sc0011941.

\bibliographystyle{JHEP}
\bibliography{refs}

\end{document}